\documentstyle[preprint,aps,epsf,floats,psfig]{revtex}

\tighten

\def\OMIT#1{{}}

\def\CA{{\cal A}}

\def\si{{}^1\kern-.14em S_0}
\def\siii{{}^3\kern-.14em S_1}
\def\piii{{}^3\kern-.14em P_1}
\def\diii{{}^3\kern-.14em D_1}

\def\pislash{ {\pi\hskip-0.6em /} }
\def\nopi{ {\rm EFT}(\pislash) }
\newcommand{\mpi}{m_\pi}
\newcommand{\mpis}{m_\pi^2}
\newcommand{\ga}{g_A}
\newcommand{\gas}{g_A^2}

\begin{document}

\preprint{\vbox{
\hbox{INT-PUB-02-48}\hbox{NT@UW-02-033}
}}

\vskip 2.0cm

\title{Partially-Quenched Nucleon-Nucleon Scattering}
\author{{\bf Silas R.~Beane}}
\address{Institute for Nuclear Theory,
University of Washington, Seattle, WA 98195-1550}

\author{{\bf Martin J.~Savage}}
\address{Department of Physics,
University of Washington, Seattle, WA 98195-1560}

\maketitle

\begin{abstract} 
  
  Nucleon-nucleon scattering is studied to next-to-leading order in a
  partially-quenched extension of an effective field theory used to describe
  multi-nucleon systems in QCD. The partially-quenched nucleon-nucleon
  amplitudes will play an important role in relating lattice simulations of the
  two-nucleon sector to nature.

\end{abstract}

\bigskip
\vskip 8.0cm
\leftline{October 2002}

\vfill\eject

\section{Introduction}

The next decade promises to be a very exciting time for strong interaction
physics. With ever increasing computer power and impressive progress in
developing new techniques to simulate quantum field theories, one hopes that
lattice simulations of simple hadronic systems will provide rigorous and
reliable predictions of QCD for strong interaction observables.  While one
looks forward to fully-unquenched QCD simulations performed with the physical
values of the light-quark masses, $m_q$, such simulations are presently
prohibitively time-consuming. At present, and for the foreseeable future,
unphysical theories will be simulated because, in contrast with QCD, the
simulations can be performed in a reasonable time frame~\cite{steverev}.  A
second motivation for simulating unphysical theories is particular to nuclear
physics, and is related to the unnaturally large values of the S-wave
scattering lengths. We will discuss this point in the final section of the
paper. A commonly simulated unphysical theory is quenched QCD (QQCD) where
disconnected quark-loop diagrams (the quark determinant) are omitted.  While
QQCD simulations of strong interaction observables can be performed with small
$m_q$, they have the distinct disadvantage of not being related to QCD except
in the large-$N_c$ limit~\cite{Chen:2002mj}. A more interesting unphysical
theory is partially-quenched QCD (PQQCD)~\cite{BG94,Sh97,GL98,SS00,SS01} in
which the quark masses, $m_S$ (the ``$S$'' stands for ``sea''), used in
evaluating the disconnected quark-loops are larger than the masses of the
quarks connected to external sources, $m_V$ (the ``V'' stands for ``valence'').
By computing strong interaction observables in PQQCD and performing the
extrapolation $m_S, m_V\rightarrow m_q$ one recovers the QCD observables that
one is interested in. It is for this last step ---the extrapolation to the
physical values of $m_q$--- that effective field theory (EFT) is required.  The
EFT's describing QCD in the low-momentum regime in the pseudo-Goldstone boson
sector (chiral perturbation theory, $\chi$PT), and in the single baryon sector
(heavy baryon chiral perturbation theory, HB$\chi$PT), are well established,
and their extension to PQQCD in the form of
PQ$\chi$PT~\cite{BG94,Sh97,GL98,SS00,SS01} and
PQHB$\chi$PT~\cite{LS96,CS01b,BS02b} have been accomplished relatively
recently.

The construction of an EFT to describe the low-momentum dynamics of
multi-nucleon systems has proven to be extremely challenging.  In the very
low-momentum regime, where the typical momentum of the external particles
involved in a given process is much less than the mass of the pion, $p\ll
m_\pi$, and hadronic production is therefore kinematically forbidden, an EFT,
$\nopi$~\cite{nopiML,nopivK,nopiCRS,Beane:2000fi}, can be constructed from
nucleons and photons (and any other low-momentum transfer probes) quite simply.
The fact that there is a bound state near threshold in the $\siii-\diii$
coupled-channels, and a pole on the second-sheet near threshold in the $\si$
channel means that at least one operator in the $\nopi$ Lagrange density must
be treated non-perturbatively in these channels.  The choice of operators to be
resummed and details of the perturbative expansion are defined by the
power-counting in $\nopi$.  Despite chiral symmetry not being a good symmetry
for $p\ll m_\pi$, isospin remains a good symmetry.  The only input into the
construction of $\nopi$ is Lorentz invariance, electromagnetic gauge
invariance, baryon number conservation and the approximate isospin symmetry,
the breaking of which can be included perturbatively.  In the kinematic regime
where the momenta involved in a given process are larger than $m_\pi$, the pion
must be included as a dynamical field. It was Weinberg's pioneering
efforts~\cite{weinberg} in the early 1990's in this kinematic regime that
initiated interest in developing EFT for nuclear physics.  Weinberg attempted
to construct an EFT for nuclear processes and nuclei involving momenta all the
way up to the chiral symmetry breaking scale $\Lambda_\chi$, and necessarily
included the pion as a dynamical degree of freedom. The power-counting that he
developed, known as Weinberg power-counting (W), involves a chiral expansion of
the nucleon-nucleon potential using the same power-counting rules that are used
in the meson and single nucleon sectors.  The chirally expanded potential is
inserted into the Schr\"odinger equation to determine observables, such as
phase shifts.  Unfortunately, there is a formal problem with this power
counting~\cite{KSW96} in some of the scattering channels, particularly the
$\si$ channel.  However, extensive phenomenological studies with W
power-counting appear to be in good agreement with data~\cite{ray,reviews,parkito,ulfito},
where such comparisons are possible, and the formal problems appear to have
little impact when a massive regulator is used with a mass scale that is not
radically different from a few hundred MeV. The formal problems with W
power-counting led Kaplan, Savage and Wise (KSW) to develop a
power-counting~\cite{KSW98} in which the momentum-independent four-nucleon 
operator is promoted to one lower order in the chiral
expansion, and consequently pion exchanges are subleading and treated in
perturbation theory.  This power-counting is formally consistent and gives
renormalization group invariant amplitudes order-by-order in the EFT expansion.
However, Fleming, Mehen and Stewart~\cite{FMS} (FMS) showed
that the scattering amplitude in the $\siii-\diii$ coupled channels diverges at
next-to-next-to-leading order
(NNLO) at relatively small momenta and KSW power-counting fails.  FMS found that
a contribution that remains large in the chiral limit destroys the convergence:
it is the chiral limit of the tensor force that ``does the damage''.  Recently,
it was suggested that one should expand observables about the chiral
limit~\cite{BBSvK} (BBSvK power-counting).  BBSvK power-counting has all the
nice features of W and KSW counting: the chiral limit of the tensor force is
resummed at leading order (LO) along with the momentum- and $m_q$-independent four-nucleon
operator in the $\siii-\diii$ coupled channels, while pions are perturbative in
the $\si$ channel, and in higher partial waves, where analytic calculations are
possible.

In recent work~\cite{BSNN}, we showed that hairpin diagrams in PQQCD give rise
to a component of the nucleon-nucleon (NN) potential that falls exponentially
at long-distance and therefore does not have the Yukawa behavior found in QCD.
Thus, measuring the long-distance behavior of the NN
potential~\cite{Richards:1990xf,Mihaly:1996ue,Stewart:1998hk,Michael:1999nq}
does not provide information about QCD unless the hairpin contribution can be
removed in a rigorous way.  While the presence of this behavior is quite
discouraging, one might focus on the behavior of S-matrix elements rather than
on the NN potential itself (for a recent survey of the status of lattice
calculations of the NN potential see Ref.\cite{Richards:2000ix}). In this work
we develop the partially-quenched EFT that describes the two-nucleon sector
using BBSvK power-counting.  In channels with higher partial waves, we give
analytic expressions for scattering amplitudes and a few characteristic
scattering volumes to next-to-leading order (NLO) in the partially-quenched
EFT. In the $\si$-channel we give analytic expressions for the scattering
amplitude, scattering length and effective range to NLO in terms of the valence
and sea quark masses.  As in QCD, the $\siii-\diii$ coupled channels in the
partially-quenched EFT are somewhat more complicated; we provide the NN
potential at NLO that is required to generate the NN phase-shifts,
$\delta_{0,2}$ and mixing-parameter, $\epsilon_1$, by solving the Schr\"odinger
equation.

\section{The Partially-Quenched EFT Calculation}

In BBSvK power-counting S-matrix elements are an expansion about the chiral
limit, where the expansion parameter, $Q$, is 
$Q\sim 1/3\sim m_\pi/\Lambda_{NN}$, 
where the constant $\Lambda_{NN} = 8 \pi f^2/( g_A^2 M_N )$ is determined by
the relative size of pion exchange, and $m_\pi$ is the physical value of the
pion mass. In the two S-wave channels, the momentum and $m_q$-independent
four-nucleon operators, along with the chiral limit of
one-pseudo-Goldstone-Boson-exchange (OPGBE), are resummed to all orders (each
iteration is the same order in $Q$) and this sum constitutes the LO scattering
amplitude. In the $\si$-channel this corresponds to KSW power-counting and the
scattering amplitude can be computed analytically, while in the $\siii-\diii$
coupled channels, the scattering amplitude must be determined numerically. In the
higher partial waves, the chiral limit of OPGBE provides the LO
contribution to the scattering amplitude as the four-nucleon operators are
suppressed by additional powers of momentum.

It is the ultra-violet (UV) behavior of the theory that requires an expansion
about the chiral limit, and in particular allows control of the very singular
diagrams as $r\rightarrow 0$ in coordinate-space. The deviations
from the chiral limit in OPGBE that formally occur at higher orders in BBSvK
power-counting are UV safe and can therefore be included at NLO without
compromising the renormalizability of the theory. The price for not including
them is that the long-distance behavior of the theory must be recovered
order-by-order in perturbation theory and convergence is somewhat
slow~\cite{BBSvK}. By contrast, the two-pseudo-Goldstone-Boson-exchange (TPGBE)
diagrams are singular away from the chiral limit, and therefore only the chiral limit can be
retained at NLO; keeping the full TPGBE introduces divergences that
cannot be renormalized at NLO. In this work we compute 
to NLO in the EFT with BBSvK power-counting.  In the S-wave channels at NLO
there are contributions from OPGBE (the full meson mass dependence is
retained), from momentum- and $m_q$-independent four-nucleon operators, from the
leading momentum-dependent four-nucleon operators ($p^2$) and from the
four-nucleon operators with a single insertion of $m_q$. In the higher partial
waves, the four-nucleon operators contribute beyond NLO and thus only OPGBE
contributes at NLO.

We will work in the isospin limit of the $SU(4|2)_L\otimes SU(4|2)_R$
PQ$\chi$PT.  This means that the $u$, $d$, $\tilde u$ and $\tilde d$ quarks in
the valence and ghost sectors are degenerate and the $j$ and $l$ quarks in the
sea sector are degenerate.  The formalism for this theory can be found in
Refs.~\cite{LS96,CS01b,BS02b} and we will not describe it here.

\subsection{Partial Waves with $L > 0$}

\begin{figure}[htb]
\centerline{{\epsfxsize=3in \epsfbox{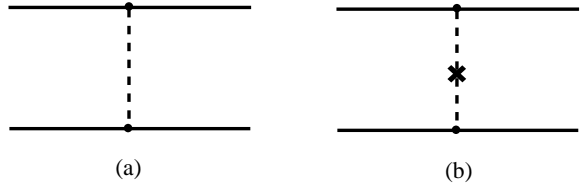}}} 
\vskip 0.15in
\noindent
\caption{
The LO contribution, ${\cal O}(Q^0)$,
to scattering in the higher partial waves from OPGBE.
Diagram (a) corresponds to the exchange of $\pi$, while
diagram (b) corresponds to the exchange of $\eta$ with
a double-pole propagator, as given in eq.~(\protect\ref{eq:props}),
denoted by ``X ''.}
\label{fig:OBE}
\vskip .0in
\end{figure}

At NLO (${\cal O}(Q^0)$), the partial waves with $L > 0$
(higher partial waves) receive contributions only
from OPGBE, as shown in Fig.~\ref{fig:OBE}, and no resummation of diagrams is
required.  In PQ$\chi$PT the potential between two-nucleons due to OPGBE
is~\cite{BSNN}
\begin{eqnarray}
V^{(PQ)} (r)\ =\ 
{1\over 8\pi f^2}\ 
{\bf\sigma}_1\cdot {\bf\nabla} {\bf\sigma}_2\cdot {\bf\nabla} 
\left( g_A^2\ {{{\bf\tau}_1\cdot{\bf\tau}_2}\over r} -
g_0^2\ {{\left(m_{SS}^2-m_\pi^2\right)}\over{2 m_\pi}}
\right) \ e^{-m_\pi r} 
\ \ ,
\end{eqnarray}
arising from the interaction Lagrange density in PQQCD 
(in the isospin limit)
\begin{eqnarray}
{\cal L} & = & 
N^\dagger\ \left[\ 
{g_A\over\sqrt{2} f} \tau^\alpha {\bf \sigma}\cdot {\bf \nabla} \pi^\alpha
\ +\ 
{g_0\over\sqrt{2} f}{\bf \sigma}\cdot {\bf \nabla}\eta\ \right] N
\ \ \ .
\end{eqnarray}
The $\pi$ and $\eta$ propagators are of the form
\begin{eqnarray}
G_{\pi} & = & {i\over q^2-m_\pi^2 + i \epsilon}
\ \ ,\ \ 
G_{\eta} \ =\  {i ( m_{SS}^2 - m_\pi^2)\over (q^2-m_\pi^2 + i \epsilon)^2}
\ \ \ ,
\label{eq:props}
\end{eqnarray}
where $f\sim 132~{\rm MeV}$, $g_A$ is the isovector axial coupling constant and
$g_0$ is the isoscalar axial coupling.  The mass $m_{SS}$ is that of a meson
composed of two sea quarks while $m_\pi$ is the pion mass which is, of course,
composed of two valence quarks. The propagators in eq.~(\ref{eq:props}) clearly
exhibit the correct behavior in the QCD limit, where the coefficient of the
double-pole contribution in the $\eta$ propagator vanishes, and the single pole
contribution is absent.  By treating $m_\pi, m_{SS}$ and $|{\bf q}|$ all of
${\cal O}(Q)$, both OPGBE contributions are the same order in the
power-counting.

In QCD, the scattering amplitudes in the higher partial waves, 
${\cal A}^{\rm (QCD)}_{J,LL^\prime, S}$, for total angular momentum $J$, total spin $S$, and
initial and final state orbital angular momentum $L$ and $L^\prime$
respectively, are well known to ${\cal O}(Q)$~\cite{EWbook}.  In the
spin-singlet channel ($S=0$) the scattering amplitude in partial waves with
$J=L>0$, ${\cal A}^{\rm (QCD)}_{J,JJ,0} $, is
\begin{eqnarray}
{\cal A}^{\rm (QCD)}_{J,JJ,0} & = &\  -\ {(-)^I\over 2I+1}\  
{3 g_A^2\over 2 f^2} \ (z-1)\  Q_L (z)
\ \ \ ,
\label{eq:ALS0}
\end{eqnarray}
where $Q_n(z)$ is an irregular Legendre Polynomial of order $n$,
using the conventions of Ref.~\cite{EWbook},
\begin{eqnarray}
Q_0(z) & = & {1\over 2}\log\left({z+1\over z-1}\right)
\ \ ,\ \ 
Q_1(z) \ =\  {1\over 2} z \log\left({z+1\over z-1}\right) -1
\ \ ,\ \ ....
\ \ \ ,
\label{eq:irregLs}
\end{eqnarray}
and where the variable $z$ is given in terms of the pion mass and nucleon
center-of-mass momentum as $z= 1 + {m_\pi^2/ ( 2 p^2) }$.
The quantity $I$ is the isospin of the channel under consideration.
In the spin-triplet channel ($S=1$) the expression is somewhat 
more complicated due to the fact that the spin and orbital angular 
momentum can  couple to produce three different total angular 
momentum states, $J=L-1, L, L+1$.
It is straightforward to show that
the amplitudes for scattering between states of the same orbital angular
momentum are~\cite{EWbook}
\begin{eqnarray}
{\cal A}_{J,LL,1}^{\rm (QCD)} & = & 
+ \ {(-)^I\over 2I+1}\ {g_A^2\over 2 f^2}
\left[\ (z-1) Q_L (z) 
\phantom{Pooh\over Pooh}
\right.\nonumber\\ && \left.\qquad\qquad\qquad
\ +\ S_{12}^{JLL} \left(\  \left(z-1\right) Q_L (z) + {3\over 2 (2L+1)} \left(
 Q_{L-1} (z) -  Q_{L+1} (z) \right)\ \right)\ \right]
\ ,
\label{eq:ALS1}
\end{eqnarray}
where the constant $S_{12}^{JLL}$ is given by
\begin{eqnarray}
S_{12}^{JLL} & = & \left(\ 
-{2(L+1)\over 2L-1}\ , \ +2\ ,\ -{2L\over 2L+3}\ \right)
\qquad {\rm for }\qquad
J\ =\ \left(\ L-1\ ,\ L\ ,\ L+1\ \right)
\ \ \ .
\end{eqnarray}
The amplitudes for scattering between states with orbital angular momenta that
differ by two units, $\Delta L=2$, induced by the tensor component of the 
interaction are
\begin{eqnarray}
{\cal A}_{J,LL+2,1}^{\rm (QCD)} & = & 
-{(-)^I\over 2I+1}\ {g_A^2\over 4 f^2}\ S_{12}^{JLL+2}\ 
\left[\ Q_{L+2} (z) +  Q_{L} (z) - 2  Q_{L+1} (z)\ \right]  
\ \ \ ,
\end{eqnarray}
where $S_{12}^{JLL+2}$ is given by 
\begin{eqnarray}
S_{12}^{JLL+2} & = & {6\sqrt{J (J+1)}\over 2 J+1}
\ \ \ .
\end{eqnarray}

In order to arrive at the partially-quenched amplitudes it is 
convenient to note that the contribution from single $\eta$ exchange 
can be obtained from OPGBE by taking a 
derivative of ${\cal A}^{\rm (QCD)}_{J,LL^\prime, S} $ with respect to 
$m_\pi^2$, and multiplying by appropriate constant factors.
Generically, 
\begin{eqnarray}
{\cal A}_{OPGBE}^{\rm (PQ)} & = & \left(\, 1\, -\, (-)^I (2I+1)\, {{g_0^2}\over{3{g_A^2}}}\,
({m_{SS}^2 - m_\pi^2\over{2p^2}})\, {\partial\over{\partial z}}\,\right)
{\cal A}_{OPGBE}^{\rm (QCD)} 
\ \ \ .
\label{eq:PQgenerator}
\end{eqnarray}
It is straightforward to show that the partially-quenched amplitudes 
in the spin-singlet higher partial waves are
\begin{eqnarray}
{\cal A}^{\rm (PQ)}_{J,JJ,0} & = &{\cal A}^{\rm (QCD)}_{J,JJ,0}
\ +\  
{g_0^2\over 4 f^2}\ {m_{SS}^2 - m_\pi^2\over  p^2} \ 
\left[\ 
 Q_L (z)\ +\ {L+1\over z+1}\left(\ Q_{L+1} (z) - z\  Q_L (z)\ \right)
\ \right]
\ ,
\label{eq:ALS0pq}
\end{eqnarray}
and in the spin-triplet higher partial waves are
\begin{eqnarray}
{\cal A}_{J,LL,1}^{\rm (PQ)} & = & 
{\cal A}^{\rm (QCD)}_{J,LL,1}
\nonumber\\
& - & 
{g_0^2\over 24 f^2}\ {m_{SS}^2 - m_\pi^2\over  p^2} \ 
\left[\ 
2\ \left(1+ S_{12}^{JLL} \right)\left( 
Q_L (z) + {L+1\over z+1}\left(\ Q_{L+1} (z) - z  Q_L (z)\ \right)\right)
\right.\nonumber\\ &&\left.
+ 
{3\  S_{12}^{JLL} \over (z^2-1)(2L+1)}
\left(L  \left[\ Q_L (z) - z  Q_{L-1} (z)\ \right]  
+ (L+2)\left[ \ z Q_{L+1} (z)
 - Q_{L+2} (z)\ \right]  \right) 
\right]
\ ,
\nonumber\\
{\cal A}_{J,LL+2,1}^{\rm (PQ)} & = & 
{\cal A}^{\rm (QCD)}_{J,LL+2,1}
\nonumber\\
& + &
{g_0^2\over 24 f^2}\ {m_{SS}^2 - m_\pi^2\over  p^2} \ 
{ S_{12}^{JLL+2} \over z^2-1}
\left[\ 
L\ \left( Q_{L+1} (z) + Q_{L+3} (z) - 2 Q_{L+2} (z) \right)
\right.\nonumber\\ && \left. 
\ -\ 
z \left( Q_{L} (z) + 3 Q_{L+2} (z) - 4 Q_{L+1} (z) \right)
\ -\ 
z L \left( Q_{L} (z) +  Q_{L+2} (z) - 2 Q_{L+1} (z) \right)
\right.\nonumber\\ && \left. \qquad \qquad \qquad \qquad
\ +\ 
 \left( Q_{L+1} (z) +  3 Q_{L+3} (z) - 4 Q_{L+2} (z) \right)
\right]
\ .
\label{eq:ALS1pq}
\end{eqnarray}

The phase-shifts in the spin-singlet channels can be easily extracted from the
scattering amplitudes given in eq.~(\ref{eq:ALS0pq}) by using the relation (for
non-relativistic systems)
\begin{eqnarray}
\delta_{J,JJ,0} & = & {1\over 2 i} \log\left( 1 + i { M_N p\over 2 \pi}
{\cal A}_{J,JJ,0}
\right)
\ \ \ ,
\label{eq:phase}
\end{eqnarray}
from which parameters in the effective range expansion can be determined for 
$p < m_\pi, m_{SS}$.  It is somewhat more complicated to determine the
phase-shifts in the spin-triplet channels as one has to disentangle them from
the mixing parameters.  However, at NLO (${\cal O}(Q^0)$), the mixing effects
are higher order in the EFT and one can straightforwardly determine parameters
in the effective range expansion.  In the P-waves, the scattering volumes,
defined to be
\begin{eqnarray}
a (^{2S+1} P _J) & = & -\lim_{p\rightarrow 0} {\tan\delta_{J,11,S}\over p^3}
\ \ \ ,
\end{eqnarray}
are found to be, at NLO,
\begin{eqnarray}
a (^1 P_1) & = & { g_A^2 M_N\over 4\pi f^2 m_\pi^2}
\ +\ 
{g_0^2 M_N\over 12\pi f^2 m_\pi^2}\ {m_{SS}^2-m_\pi^2\over m_\pi^2}
\nonumber\\
a (^3 P_0) & = & -{ g_A^2 M_N\over 4\pi f^2 m_\pi^2}
\ +\ 
{g_0^2 M_N\over 4\pi f^2 m_\pi^2}\ {m_{SS}^2-m_\pi^2\over m_\pi^2}
\nonumber\\
a (^3 P_1 ) & = & { g_A^2 M_N\over 6\pi f^2 m_\pi^2}
\ -\ 
{g_0^2 M_N\over 6\pi f^2 m_\pi^2}\ {m_{SS}^2-m_\pi^2\over m_\pi^2}
\nonumber\\
a (^3 P_2 ) & = & 0
\ \ \ ,
\end{eqnarray}
for which the QCD limit agrees with the well-known results~\cite{EWbook}.

\subsection{The $\si$ Channel}

\begin{figure}[htb]
\centerline{{\epsfxsize=3.0in \epsfbox{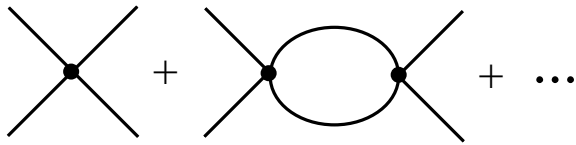}}} 
\vskip 0.15in
\noindent
\caption{
The LO contribution, ${\cal O}(Q^{-1})$,
to the scattering amplitude in the $\si$ channel.
}
\label{fig:bubbles}
\vskip .0in
\end{figure}

\begin{figure}[htb]
\centerline{{\epsfxsize=4.5in \epsfbox{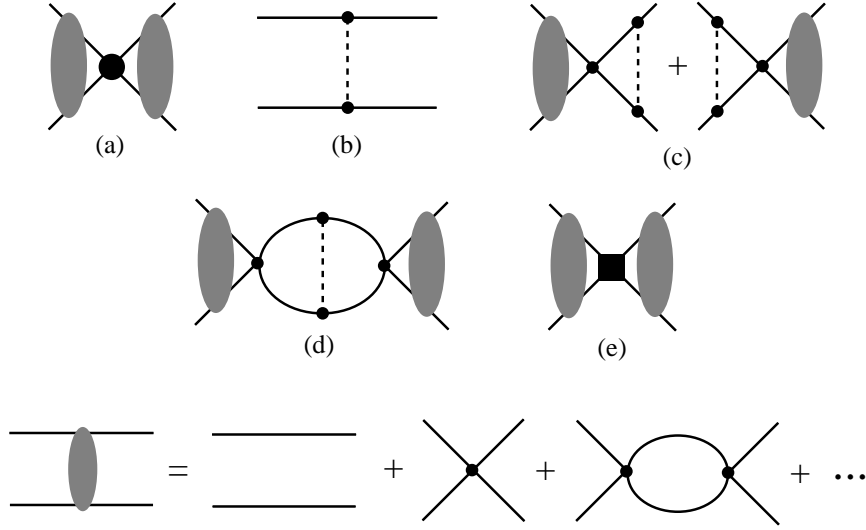}}} 
\vskip 0.15in
\noindent
\caption{
The NLO contributions, ${\cal O}(Q^{0})$,
to the scattering amplitude in the $\si$ channel.
Diagram (a) corresponds to  an insertion of the momentum-dependent operator
with coefficient $C_2^{(\siii)}$,
diagrams (b)-(d) correspond to dressed OPGBE,
while diagram (e) denotes an insertion of the $m_q$-dependent operator 
with coefficient $D^{(\si)}_2$.
}
\label{fig:nlosing}
\vskip .0in
\end{figure}

The scattering amplitude in the $\si$ channel can be determined analytically
order-by-order in perturbation theory as BBSvK power-counting coincides with
KSW power-counting in this channel.  The momentum and $m_q$-independent four
nucleon operator with coefficient $C^{(\si)}_0$ enters at LO, and the bubble
chains that it generates, as shown in Fig.~\ref{fig:bubbles}, are resummed to
all orders to produce the LO scattering amplitude~\cite{KSW98}.
At NLO there are several different contributions.  There is a contribution from
OPGBE that can be dressed in a variety of ways by the LO amplitude as shown in
Fig.~\ref{fig:nlosing}, and each dressing remains ${\cal O}(Q^0)$.  There is a
contribution from a momentum-dependent ($p^2$) operator with coefficient
$C_2^{(\si)}$ that is dressed by the LO amplitude.  Also, there are two
contributions from a single insertion of $m_q$, with coefficients
$D_{2A}^{(\si)}$ and $D_{2B}^{(\si)}$, which are also dressed by the LO
amplitude.

The scattering amplitude at NLO, $\CA^{(QCD)}_{\si}$, is the sum of the
contributions shown in Figs.~\ref{fig:bubbles} and \ref{fig:nlosing},
\begin{eqnarray}
\CA^{(QCD)}_{\si} & = & \CA^{(QCD)}_{\si , -1}
\ +\ \sum_i \CA_{\si , 0}^{(QCD)(i)}
\ \ \ \  .
\end{eqnarray}
It is straightforward to show that the individual contributions are
\begin{eqnarray}
\CA^{(QCD)}_{\si , -1} & = & 
-{ C^{(\si)}_0\over  
1 + C^{(\si)}_0 {M_N\over 4\pi}  \left( \mu + i p \right) }
\ ,
\nonumber\\
 {\cal A}^{(QCD)(I)}_{\si , 0} & = & 
-C_2^{(\si)} p^2
\left[ {\CA^{(QCD)}_{\si , -1}\over C_0^{(\si)}  } \right]^2
\ ,\ 
 {\cal A}_{\si , 0}^{(QCD)(II)}\ =\  
\left({g_A^2\over 2f^2}\right) \left(-1 + {m_\pi^2\over
4p^2} \ln \left( 1 + {4p^2\over m_\pi^2}\right)\right)
\ ,
\nonumber\\
 {\cal A}_{\si , 0}^{(QCD)(III)} &=& 
{g_A^2\over f^2} \left( {m_\pi M_N{\cal A}^{(QCD)}_{\si , -1}\over 4\pi}
\right) \Bigg( - {(\mu + ip)\over m_\pi}
+ {m_\pi\over 2p} X(p,m_\pi)\Bigg)
\ ,
\nonumber\\
{\cal A}_{\si , 0}^{(QCD)(IV)} &=& 
{g_A^2\over 2f^2} \left({m_\pi M_N{\cal A}^{(QCD)}_{\si , -1}\over
4\pi}\right)^2 \Bigg(1 -\left({\mu + ip\over m_\pi}\right)^2
+ i X(p,m_\pi)  - \ln\left({m_\pi\over\mu}\right)  
\Bigg)
\ ,
\nonumber\\
{\cal A}_{\si , 0}^{(QCD)(V)} &=& - D^{(\si)}_2 m_\pi^2 
\left[ {\CA^{(QCD)}_{\si , -1}\over C_0^{(\si)}  }\right]^2
\ ,\ 
X(p,m_\pi) \ =\  \tan^{-1} \left({2p\over m_\pi}\right) + {i\over 2} \ln
\left(1+ {4p^2\over m_\pi^2} \right)
\  ,
\label{eq:amp1s0}
\end{eqnarray}
where $D^{(\si)}_2 = D^{(\si)}_{2A} + D^{(\si)}_{2B}$, and $\mu$ is the
renormalization scale.  The PDS subtraction procedure~\cite{KSW98} has been
used in defining the power-law divergent loop diagrams.

\begin{figure}[htb]
\centerline{{\epsfxsize=3.5in \epsfbox{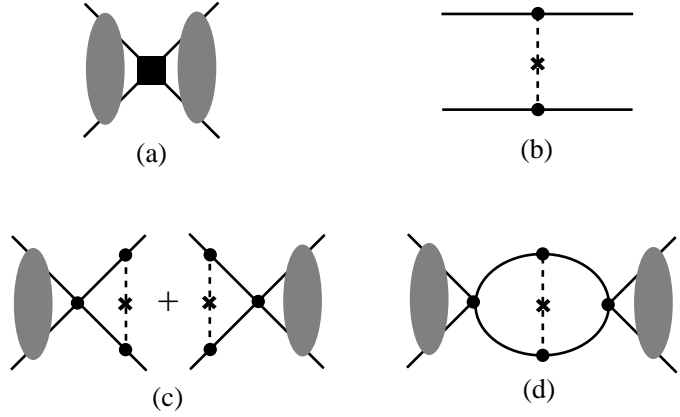}}} 
\vskip 0.15in
\noindent
\caption{
Additional contributions to the scattering amplitude in the $\si$
channel at NLO, ${\cal O}(Q^{0})$, in the partially-quenched EFT.
Diagram (a) denotes an insertion of the operator with coefficient
$D_{2B}^{(\si)}$, and diagrams (b)-(d) denote the dressed
$\eta$-exchanges.
}
\label{fig:1S0PQ}
\vskip .0in
\end{figure}

The partially-quenched amplitude in the $\si$ channel can be found
straightforwardly from the QCD amplitude by taking derivatives with respect to
$m_\pi^2$ of the OPGBE contributions (see eq.~(\ref{eq:PQgenerator}))
and by constructing the local operators
that can contribute in the $SU(4|2)_L\otimes SU(4|2)_R$ EFT.  The additional
diagrams that contribute are shown in Fig.~\ref{fig:1S0PQ},
and the scattering amplitude in the $\si$ channel at NLO is 
\begin{eqnarray}
\CA^{(PQ)}_{\si} & = & \CA^{(QCD)}_{\si}
\ -\ 
\left(m_{SS}^2-m_\pi^2 \right) 
\left({\CA^{(QCD)}_{\si , -1}\over C^{(\si)}_0 }\right)^2\
D_{2B}^{(\si)} (\mu)
\nonumber\\
& + &  
{g_0^2\over 2 f^2}\ {m_{SS}^2-m_\pi^2\over 2 p^2}\ 
\left[\ 
{1\over 2} \log\left(1 + {4p^2\over m_\pi^2}\right)
\ -\ {2p^2\over m_\pi^2+4p^2}
\right.\nonumber\\ & & \left.\qquad
\ +\ i {M_N p\over 2\pi}\CA^{(QCD)}_{\si , -1} 
\left(1 + i {M_N p\over 4\pi} \CA^{(QCD)}_{\si , -1}\right)
\left( \log\left(1 - i {2p\over m_\pi}\right) + i {p\over m_\pi-i 2 p}\right)
\right.\nonumber\\ & & \left.\qquad
 - {M_N^2p^2\over 8\pi^2} \left(\CA^{(QCD)}_{\si , -1}\right)^2 
\left(\ \log\left({m_\pi\over\mu}\right)\ -\ {1\over 2}\ \right)
\right]
\ \ \ \  ,
\end{eqnarray}
where we have worked to LO in the relation between the quark masses and
the meson masses.  The explicit renormalization-scale dependence of the
amplitude due to the $\log\left({m_\pi\over\mu}\right)$ contribution is exactly
compensated by the renormalization-scale dependence of the coefficient
$D_{2B}^{(\si)} (\mu)$ to yield a $\mu$-independent amplitude.

As there is no mixing between different partial waves in the $\si$ channel it
is straightforward to determine the phase-shift, $\delta_{\si} =
\delta_{0,00,0}$, from the scattering amplitude without approximation using
eq.~(\ref{eq:phase}), and, in turn, to construct the effective range expansion
\begin{eqnarray}
p \cot\delta_{\si} - i p & = & {4\pi\over M_N} {1\over\CA_{\si}}
\ =\ -{1\over a^{(\si)}} + {1\over 2} r^{(\si)} p^2 + ...
\ \ \ \  .
\end{eqnarray}
Here $a^{(\si)}$ and $r^{(\si)}$ are the scattering length and effective range
in the $\si$ channel, respectively.  The scattering length in the
partially-quenched EFT is found to be
\begin{eqnarray}
{1\over a^{(\si)}} & = & \gamma
\ -\ {M_N\over 4\pi}\  (\mu-\gamma)^2\ D_2^{(\si)} (\mu)\ m_\pi^2
\ -\ {M_N\over 4\pi}\  (\mu-\gamma)^2\ D_{2B}^{(\si)} (\mu)\ 
\left(m_{SS}^2-m_\pi^2\right)
\nonumber\\
&&\ +\ 
{g_A^2 M_N\over 8\pi f^2}\left[\ 
m_\pi^2\log\left({\mu\over m_\pi}\right) \ +\ 
(m_\pi-\gamma)^2 - (\mu-\gamma)^2\ \right]
\nonumber\\
&&\ +\ 
{g_0^2 M_N\over 8\pi f^2}\ 
\left(\ m_{SS}^2-m_\pi^2\ \right)\ 
\left[\ \log\left({\mu\over m_\pi}\right) \ +\ 
{1\over 2}  - {\gamma \over m_\pi}\ \right]
\ \ \ \  ,
\end{eqnarray}
where $\gamma$ is a $\mu$-independent linear combination of $C_0^{(\si)}$ and $\mu$ 
that enters at LO in the expansion and must be determined from data.
Furthermore, the effective range is found to be
\begin{eqnarray}
r^{(\si)} & = & {M_N\over 2\pi} (\mu-\gamma)^2 C_2(\mu)
\ +\ 
{g_A^2 M_N\over 12\pi f^2}\ \left( 3 - 8{\gamma\over m_\pi} + 6 {\gamma^2\over
    m_\pi^2}\ \right)
\nonumber\\&&
+\ 
{g_0^2 M_N\over 6\pi f^2} \ {m_{SS}^2-m_\pi^2\over m_\pi^2}\ 
\left(\  2 {\gamma\over m_\pi} - 3 {\gamma^2\over m_\pi^2}\ \right)
\ .
\end{eqnarray}

In QCD with KSW power-counting, the scattering amplitude in the $\si$ channel
has been determined up to NNLO~\cite{FMS}, and it has been found that the
expansion is convergent.  However, the chiral expansion of the effective range
parameters in this channel suggests that the convergence of the expansion is
quite slow~\cite{CoHa99}. Consequently, in order to have confidence
in the chiral extrapolation of the partially-quenched amplitude and effective
range parameters we have presented here, the NNLO amplitude (and even higher
orders) should be computed in order to understand the convergence properties of
the chiral expansion.

\subsection{The $\siii-\diii$ Coupled Channels}

\begin{figure}[!ht]
\centerline{{\epsfxsize=3.63in \epsfbox{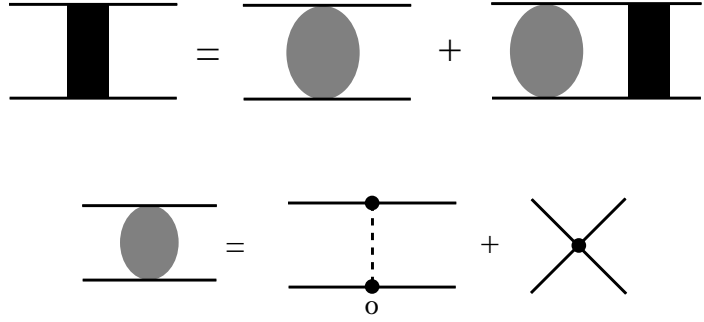}} }
\vskip 0.15in
\noindent
\caption{Lippmann-Schwinger equation for the LO contribution to the
  scattering amplitude (large solid rectangle) in the 
$\siii -\diii$ coupled-channels. 
The small solid circles denote an insertion of 
$C_0^{(\siii)}$ or $g_A$. The ``o'' appearing below the
OPGBE diagram implies the chiral limit.}
\label{fig:3S1LO}
\vskip .2in
\end{figure}

\begin{figure}[!ht]
\centerline{{\epsfxsize=5.5in \epsfbox{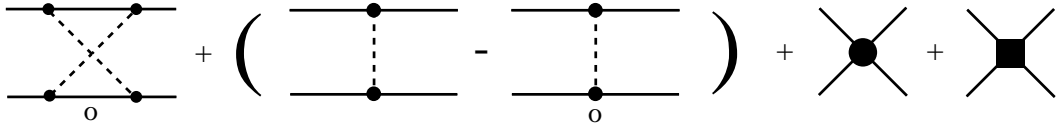}} }
\vskip 0.15in
\noindent
\caption{Chiral limit of the crossed TPGBE diagram, deviations from the chiral
  limit of OPGBE, and the $C_2$
  (large solid circle) and $D_2$ (large solid square) operators, all of which
  contribute at NLO in the $\siii -\diii$ coupled-channels. The ``o''
  appearing below a diagram implies the chiral limit.}
\label{fig:3S1NLO}
\vskip .2in
\end{figure}

Due to the non-perturbative nature of OPGBE ---particularly the chiral limit of
the tensor force--- in the $\siii-\diii$ channel, the method for computation in
this channel is fundamentally different from that in the $\si$ channel and the
higher partial waves where OPGBE can be included in perturbation theory.  The
details of the calculation of scattering lengths, phase shifts and bound state
energies in the $\siii-\diii$ coupled channels in QCD can be found in
Refs.~\cite{BBSvK,BSmq}, and we do not repeat them here.  In the $\siii-\diii$
coupled-channels, OPGBE generates both central and tensor potentials,
\begin{eqnarray}
V_C^{(QCD)(\pi)} (r;\mpi) & = & -{\alpha_\pi}\,\mpis\;{e^{-\mpi r}\over r} 
\nonumber\\
V_T^{(QCD)(\pi)} (r;\mpi) &=& -{\alpha_\pi}\ {e^{-\mpi r}\over r}
\left(\ {3\over r^2} + {3 m_\pi \over r}\ +\ m_\pi^2\ \right)
\ \ ,
\label{eq:OPEpots}
\end{eqnarray}
where ${\alpha_\pi}= \gas(1-2m_\pi^2 \overline{d}_{18}/\ga)^2/(8\pi f^2)$. 
The constant $\overline{d}_{18}$ is somewhat uncertain~\cite{ulfioffe},
with different extractions yielding $-0.78\pm 0.27$, $-0.83\pm 0.06$, 
$-1.4\pm 0.24$~\cite{Fettes:1999wp} and 
$-10.14\pm 0.45~{\rm GeV}^{-2}$~\cite{Fettes:2001cr}.
Since the chiral limit of the potentials in eq.~(\ref{eq:OPEpots})
contribute at LO, as shown in Fig.~\ref{fig:3S1LO}, the $m_q$-dependence of
$g_A$, $f$ and $M_N$ are required at NLO~\cite{BBSvK}.
Each of these observables has been studied extensively, the results of which
can be found in Refs.~\cite{ulfioffe,GaLe84,CoGaLe01,FeMe00}, and up to 
NNLO it is found that
\begin{eqnarray}
f & = & f^{(0)} \left[ 1 
 - {1 \over 4\pi^2 (f^{(0)})^2 } 
m_\pi^2\log\left({m_\pi\over m_\pi^{({\rm PHYS})}}\right)
+ { m_\pi^2\over 8\pi^2 (f^{(0)})^2} 
\overline{l}_4  \right]
\nonumber\\
M_N & = & M_N^{(0)} - 4 m_\pi^2 c_1 
\nonumber\\
g_A & = & g_A^{(0)} 
\left[ 1 - { 2 (g_A^{(0)})^2+1
\over 4\pi^2 (f^{(0)})^2 } 
m_\pi^2\log\left({m_\pi\over m_\pi^{({\rm PHYS})}}\right)
- { (g_A^{(0)})^2 m_\pi^2 \over 8\pi^2 (f^{(0)})^2}
+ {4 m_\pi^2\over g_A^{(0)} } \overline{d}_{16}
\right]
\ \ \ ,
\label{eq:SNparams}
\end{eqnarray}
where $m_\pi^{({\rm PHYS})}=139~{\rm  MeV}$,
$\overline{l}_4=4.4\pm 0.2$~\cite{GaLe84,CoGaLe01}, 
$c_1 \sim -1~{\rm  GeV}^{-1}$~\cite{ulfioffe} are $m_q$-independent constants
(we have explicitly separated the logarithmic contribution from 
$\overline{l}_4$).
A complete analysis by Fettes~\cite{Fettes:fd}
of the $\pi N$ sector
provides three determinations of $\overline{d}_{16}$, 
$\overline{d}_{16}=-0.91\pm 0.74$, $-1.01\pm 0.72$ and 
$-1.76\pm 0.85~{\rm GeV}^{-2}$.
At NLO there is a contribution from the chiral limit of TPGBE
and from an insertion of a momentum dependent ($p^2$) operator with coefficient
$C_2^{(\siii)}$, as shown in Fig.~\ref{fig:3S1NLO}.  At this order there are
two contributions arising from a single insertion of $m_q$, with coefficients
$D_{2A}^{(\siii)}$ and $D_{2B}^{(\siii)}$, which in QCD are combined together
into $D_2^{(\siii)}$.  The TPGBE potential in coordinate space has been computed
in Ref.~\cite{ray,KBW}, and in the chiral limit is given by
\begin{eqnarray}
V_C^{(QCD)(\pi\pi)} (r; 0) 
& = & {3(22 g_A^4- 10 g_A^2-1)\over 64\pi^3 f_\pi^4} 
{1\over r^5}
\quad , \quad
V_T^{(QCD)(\pi\pi)} (r; 0)  \ =\ -{15 g_A^4\over 64\pi^3 f_\pi^4} {1\over r^5}
\ \ .
\label{eq:TPEpots}
\end{eqnarray}

The singular nature of the tensor force has so far precluded regularization of
this channel using dimensional regularization, or any other mass-independent
regulator. We therefore use a spatial square-well of radius
$R$~\cite{BBSvK,BSmq,Sprung}, where the potential outside the square well,
${\cal V}_L^{(QCD)}(r;\mpi) = {\cal V}_L^{(QCD)}$, is
\begin{eqnarray}
{\cal V}_L^{(QCD)} & = & M_N \left(
\begin{array}{cc}   
  -V_C^{(QCD)}(r;\mpi)  & -2\sqrt{2} \; V_T^{(QCD)}(r;\mpi)\\               
-2\sqrt{2}\;  V_T^{(QCD)}(r;\mpi) & 
-  V_C^{(QCD)}(r;\mpi)+ 2 V_T^{(QCD)}(r;\mpi)-{6\over M_N r^2}
\end{array}
\right)
\ ,
\label{eq:potout}
\end{eqnarray}
where 
\begin{eqnarray}
V_C^{(QCD)} (r; m_\pi) & = & V_C^{(QCD)(\pi)} (r; m_\pi)\ +\ 
V_C^{(QCD)(\pi\pi)} (r; 0)
\nonumber\\
V_T^{(QCD)} (r; m_\pi) & = &   V_T^{(QCD)(\pi)} (r; m_\pi)\ +\ 
V_T^{(QCD)(\pi\pi)} (r; 0)
\ \ .
\label{eq:potsum}
\end{eqnarray}
The energy and $m_q$-dependent potential, 
${\cal V}_S^{(QCD)}(r;m_\pi ,k^2)={\cal V}_S^{(QCD)}$, inside the square well is
\begin{eqnarray}
{\cal V}_S^{(QCD)} & = & -M_N
\left(
\begin{array}{cc}   
V_{\rm sq}^{(QCD)}& 0\\               
0     & V_{\rm sq}^{(QCD)}+ {6\over M_N r^2}
\end{array}
\right) 
\label{eq:potin}
\end{eqnarray}
where 
$V_{\rm sq}^{(QCD)}=V_{C_0^{(\siii)}} + m_\pi^2\ V_{D_2^{(\siii)}} + p^2 V_{C_2^{(\siii)}}$ 
and where we have again used the LO relation between $m_q$
and the pion mass.  $V_{C_0^{(\siii)}}$, $V_{D_2^{(\siii)}}$ and
$V_{C_2^{(\siii)}}$ are constant potentials corresponding to the renormalized
local operators with coefficients $C_0^{(\siii)}$, $D_2^{(\siii)}$ and
$C_2^{(\siii)}$ in the $\siii-\diii$ coupled-channels, respectively.  An
identification can be made between the coefficients of the local operators,
$C_i^{(\siii)}$ and $D_i^{(\siii)}$, and the constant potentials of the
square-wells that enter into eq.~(\ref{eq:potin}), $V_{C_i^{(\siii)}}$ and
$V_{D_i^{(\siii)}}$. For example
\begin{eqnarray} 
{C_i^{(\siii)}}\ \delta^{(3)} (r) & \rightarrow & 
{{3{C_i^{(\siii)}}\  \theta (R-r)}\over 4\pi R^3}\ 
\equiv\  {V_{C_i^{(\siii)}}}\ \theta (R-r)  
\ \ \ .
\label{eq:ctovsinglet}
\end{eqnarray}
It is important to recall that there is implicit $m_q$-dependence in this
potential arising from the chiral expansions of $g_A$, $M_N$ and $f$, in
addition to the explicit dependence from the $D_2^{(\siii)} m_\pi^2$
contribution.  Defining the two-component wavefunction $\Psi$ to be
\begin{eqnarray}
\Psi (r) & = & \left(\begin{array}{c}
            u(r)\\
            w(r)
            \end{array}\right)
          \ \ \ , 
\end{eqnarray} 
where $u(r)$ is the S-wave wavefunction and $w(r)$ is the D-wave wavefunction,
the regulated Schr\"odinger equation is 
\begin{eqnarray}
{\Psi'' (r)}
\ +\  \left[\  p^2 \ +\  {\cal V}_L^{(QCD)} (r;\mpi)\ \theta (r-R) 
\ +\ {\cal V}_S^{(QCD)} (r;m_\pi ,k^2)\ \theta (R-r)\ \right] \Psi (r) & = & 0
\ \ .
\label{eq:SE} 
\end{eqnarray} 

At LO in the partially-quenched theory one has contributions to the NN
potential from the momentum and $m_q$-independent four-nucleon operator with
coefficient $C_0^{(\siii)}$ and from the chiral limit of OPGBE.  In fact, as
discussed earlier, it is consistent to retain the full $m_q$ dependence of
OPGBE potential at NLO, which in PQQCD leads to
\begin{eqnarray}
V_C^{(PQ)(\pi)} (r) & = & 
-\alpha_\pi\ m_\pi^2\ {e^{-m_\pi r }\over r}
\ -\ {\alpha_0\over 6}\ \left( m_{SS}^2-m_\pi^2\right) 
\left[ m_\pi-{2\over r}\right] e^{-m_\pi r }
\nonumber\\
V_T^{(PQ)(\pi)} (r) & = & 
-\alpha_\pi\ {e^{-m_\pi r }\over r}
\ \left( {3\over r^2} + 3 {m_\pi\over r} + m_\pi^2\right)
\ -\ {\alpha_0\over 6}\ \left( m_{SS}^2-m_\pi^2\right) 
\left[ m_\pi+{1\over r}\right] e^{-m_\pi r }
\ ,
\label{eq:PQpot}
\end{eqnarray}
where it is implicit that both $\alpha_\pi$ and $\alpha_0$ are defined in the
partially-quenched theory, whose chiral expansions differ from those of QCD.
Given that only the tensor component of OPGBE contributes at LO, and $\alpha_0$
contributes only at NLO, only the partially-quenched expansion of $\alpha_\pi$
is required for this NLO calculation.  While one can straightforwardly
construct all the operators that contribute to the process in PQ$\chi$PT as one
does in $\chi$PT, this is a tedious procedure.  Ultimately one arrives at the
following relations appropriate for an NLO calculation
\begin{eqnarray}
\alpha_\pi & = & {g_A^2\over 8\pi f^2}
\left( 1 - {2 m_\pi^2 \overline{d}_{18}\over g_A}
 - {2 (m_{SS}^2 - m_\pi^2) \overline{d}_{18B}\over g_A} 
\right)^2
\ \ ,\ \ 
\alpha_0 \ =\ {g_0^2\over 8\pi f^2}
\ \ \  ,
\end{eqnarray}
where $\overline{d}_{18B}$ is an additional coefficient 
that must be determined from lattice calculations.
In the chiral expansion of the OPGBE potentials in eq.~(\ref{eq:PQpot})
it is important to note that there is no contribution of the form 
$1/r^2$.  The appearance of such a term would destroy the renormalization
program we have constructed for QCD and it is encouraging that 
such a term does not arise from OPGBE in PQQCD either.

The chiral expansions of $f$, $M_N$ and $g_A$ are also required, and these are
known. The chiral expansion of $f$ at NLO is~\cite{Golterman:1997st}
\begin{eqnarray}
f & = & f^{(0)}\ \left[\ 
1 - {m_{SV}^2\over 4\pi^2 (f^{(0)})^2} \log\left({m_{SV}\over\mu}\right)
\ +\ l_1\ m_\pi^2\ +\ l_2\ m_{SS}^2\ \right]
\ \ \ \  ,
\end{eqnarray}
where $m_{SV}$ is the mass of a meson composed of one valence and one sea
quark, at LO in the chiral expansion, and $l_{1,2}$ are coefficients that need
to be determined from lattice calculations and are directly related to the
constant $\overline{l}_4$ in QCD.  At NLO, the nucleon mass receives
contributions from counterterms only
\begin{eqnarray}
M_N & = & M_0\ +\ c_1 m_\pi^2\ +\ c_2 m_{SS}^2\ +\ ...
\ \ \ \  ,
\end{eqnarray}
where the $c_{1,2}$ are to be determined from lattice calculations.  The matrix
element of the axial current is somewhat more complicated, as it depends upon
how one extends the axial currents from QCD to PQQCD, as discussed in
Refs.~\cite{GP01a,CS01b,BS02b}. For vanishing ghost and sea-quark axial charge
$y^{(S)}$, the axial matrix element is found to be~\cite{BS02b}
\begin{eqnarray}
g_A & = & g_A^{(0)}\ +\ 
{1\over 16\pi^2 f^2}\left(\eta^{(0)}\ -\ g_A^{(0)} w_N\ \right)\ +\ 
 c^{(0)}\ ,
\end{eqnarray}
where
\begin{eqnarray}
\eta^{(0)} & = & {({g_A^{(0)})^3}\over 2} L_\pi - 2{g_A^{(0)}} L_{SV} -
\ +\ 
{g_0^{(0)}- g_A^{(0)}\over 6}
\left( 3 (g_A^{(0)})^2  + (g_0^{(0)})^2\right)
\left( L_\pi-L_{SV}\right)
\nonumber\\
w & = & (g_A^{(0)})^2 \left( {1\over 2} L_\pi + 4 L_{SV} \right)
- 
{g_0^{(0)}- g_A^{(0)}\over 2} 
\left( 5 g_0^{(0)} - g_A^{(0)}\right) \left( L_\pi-L_{SV}\right)
\nonumber\\
c^{(0)} & = & r_1\  m_\pi^2 + r_2 \ m_{SS}^2
\ \ \   ,
\end{eqnarray}
and $L_\pi = m_\pi^2 \log\left(m_\pi^2/\mu^2\right)$, and 
$L_{SV} = m_{SV}^2 \log\left(m_{SV}^2/\mu^2\right)$.  Generalization to arbitrary ghost and
sea-quark axial charges is straightforward~\cite{GP01a,BS02b}.  The $r_i$ are
constants that must be determined from the lattice.  Unlike in the single
nucleon sector~\cite{BS02b}, we have not included the $\Delta$ resonance as an
explicit degree of freedom, as the $\Delta$-nucleon mass splitting is
approximately the same as $\Lambda_{NN}$, the scale at which the EFT is
expected to break-down. If the range of the EFT is somehow extended to higher
momenta (or if this one is shown to be valid at higher momenta that we
presently expect), the $\Delta$ should be included in the theory explicitly, as
the $\Delta N\pi$ intermediate states ---in contrast with the $\Delta\Delta$
states--- are likely to make a sizable contribution that cannot be described by
local counterterms at momenta beyond $\Lambda_{NN}$.

\begin{figure}[!ht]
\centerline{{\epsfxsize=2.5in \epsfbox{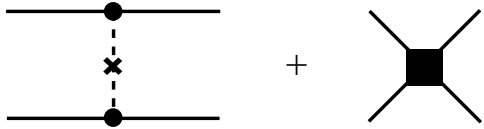}} }
\vskip 0.15in
\noindent
\caption{
Additional contributions to the scattering amplitude in the $\siii$
channel at NLO, ${\cal O}(Q^{0})$, in the partially-quenched EFT.
The large solid square denotes an insertion of an $m_q$-dependent operator with coefficient
$D_{2B}^{(\si)}$.}
\label{fig:3S1pq}
\vskip .2in
\end{figure}
At NLO in the partially-quenched EFT there are potentially additional
contributions from TPGBE.  However, it is only the chiral limit of
TPGBE that contributes at NLO and the chiral limit of PQQCD is the same as the
chiral limit of QCD, by construction, and thus the TPGBE potentials are those
given in eq.~(\ref{eq:TPEpots})~\footnote{In W counting one keeps the full
  TPGBE contribution at NLO~\cite{ray}, for which partial-quenching is
  somewhat more complicated~\cite{BSNN}.}.  Thus, the additional contributions at NLO
have the same form as in the $\si$ channel: a single insertion of $m_q$
and the exchange of a single $\eta$, as shown in Fig.~\ref{fig:3S1pq},
which generates the potentials given in eq~(\ref{eq:PQpot}).  The additional
contribution from a single insertion of $m_q$ leads to a modification of the
short-distance potential in eq.~(\ref{eq:potin}).  The short distance potential
in the partially-quenched theory is
\begin{eqnarray}
V_{\rm sq}^{(PQ)} & = & V_{\rm sq}^{(QCD)}
\ +\  \left( m_{SS}^2-m_\pi^2\right) V_{D_{2B}^{(\siii)}}
\ \ \ \  .
\end{eqnarray}

This completes the construction of a partially-quenched EFT describing the
$\siii-\diii$ coupled channels. The potential defined above is inserted into
the Schr\"odinger equation in eq.~(\ref{eq:SE}) to generate observables. Due
to the lack of partially-quenched lattice data in this channel we have not
generated phase-shifts or scattering lengths.  However, this is a
straightforward procedure and requires only limited numerical
work~\cite{BBSvK,BSmq}.

\section{Discussion}

In this work we have formulated the effective field theory required to describe
the low-energy behavior of partially-quenched QCD in the two-nucleon sector.
In fact, it is quite simple to construct the partially-quenched effective field
theory from the known QCD results and it is gratifying to see that one can
obtain analytic results for many observables in the $\si$ channel and in the
higher partial waves.  While a numerical solution is required in the
$\siii-\diii$ coupled channels, it amounts to a simple problem in
non-relativistic quantum mechanics.

One should be concerned about the range of sea and valence quark masses for
which this theory converges.  In QCD it is found that the NN EFT converges for
$m_\pi$ and momenta less than of order $\Lambda_{NN}\sim 300~{\rm MeV}$, and
one suspects that the same radius of convergence will exist in the
partially-quenched theory. If this is indeed the case, lattice calculations
will be required with meson masses of less than $\sim 300~{\rm MeV}$ in order
to match to the EFT and use it to make predictions about nature.  This is
somewhat more restrictive than in the meson and single nucleon sectors and
therefore one would like to see convergent results in those sectors before
being confident in results obtained in the multi-nucleon sectors.
\begin{figure}[!tbp]
\centerline{\psrotatefirst
\psfig{file=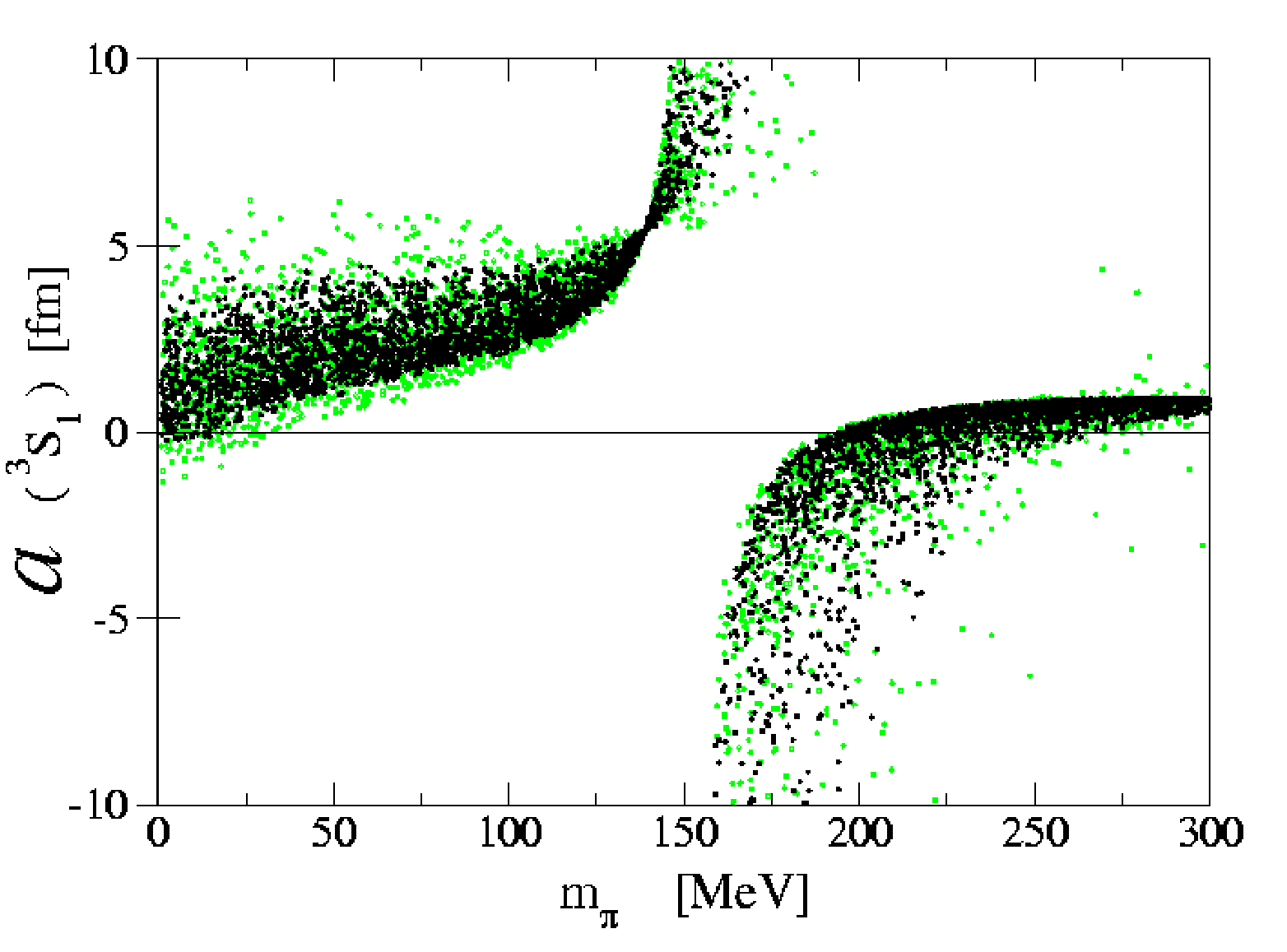,width=3.3in}}
\vskip 0.15in
\noindent
\caption{The scattering length in the $\siii$-channel in QCD as a function of the 
pion mass for characteristic strong-interaction parameters. The two shaded
regions correspond to different allowed ranges of $D^{(\siii)}_2$ that
are both consistent with naive dimensional analysis.
For a detailed discussion, see Ref.~\protect\cite{BSmq}. 
At the physical value of the pion mass the scattering length is
$a^{(\siii)}\sim +5.425~{\rm fm}$.}
\label{fig:3S1mqA}
\vskip .2in
\end{figure}

It is very encouraging to see that partially-quenched calculations of
quantities in PQ$\chi$PT describing the dynamics of the PGB's are presently
being performed~\cite{pqqcdLi}, and linear combinations of the Gasser-Leutwyler
coefficients appearing at ${\cal O}(p^4)$ in $\chi$PT are being determined as a
result. The situation is far less advanced in the two-nucleon sector.  At
present, NN scattering lengths must be extracted at finite volume using
L\"uscher's formula, which expresses the energy of a two-particle state as a
perturbative expansion in the scattering length divided by the size of the
box~\cite{Luscher}. There are (at least) two potential problems with this
approach.  First, one may worry that lack of unitarity in PQQCD may invalidate
L\"uscher's formula for the NN scattering lengths.  However, L\"uscher's
formula is easily obtained in $\nopi$~\cite{baal}, and one can convince oneself
using the arguments of Ref.~\cite{BSNN} that, while the pionful NN EFT
described above is not unitary in PQQCD, $\nopi$ is unitary in PQQCD; all the
effects of partial-quenching in $\nopi$ are in the coefficients of the contact
operators.  It follows by continuity that the NN scattering lengths can be
extracted in a lattice simulation of PQQCD by using L\"uscher's formula and
extrapolating to the physical quark masses using the formalism presented in this
paper. A second worry is truly cause for concern: the S-wave NN scattering
lengths are extremely large (which is understood as proximity to an infrared
fixed point~\cite{KSW98,birse}) as compared to the sizes of state-of-the-art lattices:
$a^{(\si)}\sim -23.714~{\rm fm}$ and $a^{(\siii)}\sim +5.425~{\rm fm}$. A recent
study~\cite{BSmq} of the pion-mass dependence of the NN scattering lengths in
QCD suggest that the $\siii$ scattering length relaxes to
\begin{figure}[ht]
\centerline{\psrotatefirst
\psfig{file=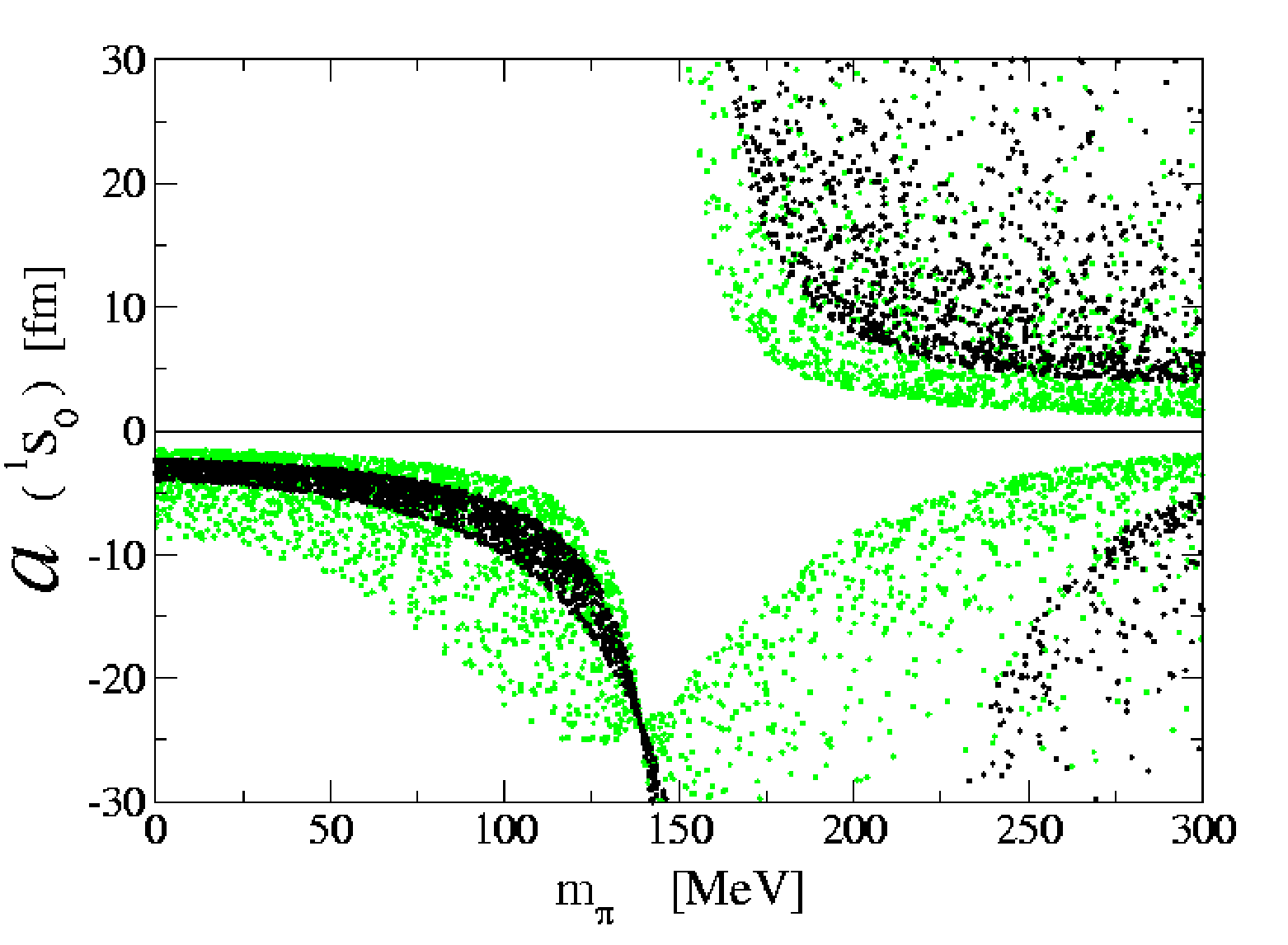,width=3.3in}}
\vskip 0.15in
\noindent
\caption{The scattering length in the $\si$-channel in QCD as a function of the 
pion mass for characteristic strong-interaction parameters. The two shaded
regions correspond to different allowed ranges of $D^{(\si)}_2$ that
are both consistent with naive dimensional analysis.
For a detailed discussion, see Ref.~\protect\cite{BSmq}. 
At the physical value of the pion mass the scattering length is
$a^{(\si)}=-23.714\pm 0.013~{\rm fm}$.
}
\label{fig:1S0mq}
\vskip .2in
\end{figure}
natural values of $\sim 1~{\rm fm}$ as the pion mass is increased beyond 
$\sim 200~{\rm MeV}$ (see Fig.~\ref{fig:3S1mqA}). One anticipates similar
behavior in the partially-quenched theory.
Given current uncertainties in
strong interaction parameters, particularly $D^{(\si)}_2$, it is at present
unclear whether the same is true in the $\si$ channel (see Fig.~\ref{fig:1S0mq}). 

To our knowledge, a single lattice determination of the $\si$ and $\siii$ NN
scattering lengths in QQCD exists~\cite{fuku} at a pion mass of $\sim 500~{\rm
  MeV}$.  This pion mass is beyond the range of applicability of the EFT
described in this paper, or of an analogous EFT that one can easily construct
to describe QQCD.  Given the unknown $D_2$ coefficients that encode the
short-distance quark-mass dependence in the S-wave channels one may question
the motivation for an improved, partially-quenched simulation of the NN
scattering lengths with pion masses within the NN EFT, since such simulation
will, at best, simply determine the $D_2$ operators. It has recently been
shown~\cite{BBSvK,BSmq,Epelbaum:2002gb} that to leading order in the NN EFT,
and assuming perfect knowledge of the single-nucleon sector, the $D_2$
operators determine the quark-mass dependence of the deuteron and, more
generally, of dinucleon binding.  Since small changes in $m_q$ can in principle lead
to drastic changes in the positions of nuclear energy levels, much attention
has been given to light-element abundances predicted by big-bang
nucleonsynthesis and to the abundance of isotopes produced by the Oklo
``natural reactor'' in the hope that these abundances can be used to constrain
high-energy physics~\cite{Uzan:2002vq}.  This is powerful motivation indeed for
improved lattice simulations of NN scattering lengths. We look forward to a
significant lattice effort in the multi-nucleon sector.

\bigskip\bigskip

\acknowledgements

This work is supported in part by the U.S. Department of Energy under 
Grant No.~DE-FG03-00-ER-41132 (SRB) and Grant No.~DE-FG03-97-ER-4014 (MJS).

\end{document}